\documentstyle[aps,epsfig]{revtex}
\begin{document}
\title{  $J/\psi$ suppression in Pb+Pb collisions, a conventional
description.}

\author{\bf A. K. Chaudhuri\cite{byline}}
\address{ Variable Energy Cyclotron Centre\\
1/AF,Bidhan Nagar, Calcutta - 700 064\\}
\date{\today}

\maketitle

\begin{abstract}

We  have analyzed the latest NA50 data on $J/\psi$ suppression in
Pb + Pb collisions. $J/\psi$ production is assumed to  be  a  two
step   process,  (i)  formation  of  $c\bar{c}$  pair,  which  is
accurately calculable in QCD and (ii) formation of $J/\psi$ meson
from   the   $c\bar{c}$   pair,   which   can   be   conveniently
parameterized.  In  a pA/AA collision, the as the $c\bar{c}$ pair
pass through the nuclear medium, it gain relative square momentum
at the rate of $\varepsilon^2$ per unit path length. As a result,
some of the $c\bar{c}$ pairs can gain enough  momentum  to  cross
the   threshold  to  become  an  open  charm  meson,  leading  to
suppression in pA/AA collisions. The parameters of the model were
fixed from experimental data on the total $J/\psi$ cross  section
as  a function of effective nuclear length. The model without any
free parameter, give excellent description of NA50 data on  $E_T$
dependence  of $J/\psi$ to Drell-Yan ratio. The model was applied
to predict the $E_T$ dependence of $J/\psi$ at RHIC energy.  Much
larger  suppression  of  $J/\psi$,  in agreement with other model
calculations are predicted. \end{abstract}


\section{Introduction}

In  relativistic  heavy  ion  collisions $J/\psi$ suppression has
been recognized as an important tool  to  identify  the  possible
phase transition to quark-gluon plasma. Because of the large mass
of  the  charm  quarks,  $c\bar{c}$ pairs are produced on a short
time scale. Their tight binding also makes them immune  to  final
state interactions. Their evolution probes the state of matter in
the  early  stage  of the collisions. Matsui and Satz \cite{ma86}
predicted that in presence of quark-gluon plasma  (QGP),  binding
of  $c\bar{c}$  pairs  into  a  $J/\psi$  meson will be hindered,
leading to the  so  called  $J/\psi$  suppression  in  heavy  ion
collisions  \cite{ma86}.  Over  the  years  several  groups  have
measured the $J/\psi$ yield in heavy ion collisions (for a review
of the data and the interpretations see Refs.  \cite{vo99,ge99}).
In  brief,  experimental  data  do show suppression. However this
could be attributed to the conventional nuclear absorption,  also
present in $pA$ collisions.

The latest data obtained by the NA50 collaboration \cite{na50} on
J/$\psi$ production in Pb+Pb collisions at 158 A GeV is the first
indication  of the anomalous mechanism of charmonium suppression,
which goes beyond  the  conventional  suppression  in  a  nuclear
environment.  The  ratio  of  J/$\psi$ yield to that of Drell-Yan
pairs decreases faster with $E_T$ in the most central  collisions
than  in  the  less  central ones. It has been suggested that the
resulting pattern can be understood in a  deconfinement  scenario
in  terms  of  successive  melting  of  charmonium  bound  states
\cite{na50}. In a recent paper , Blaizot {\em et al.} \cite{bl00}
have shown that the data  can  be  understood  as  an  effect  of
transverse  energy  fluctuations in central heavy ion collisions.
Introducing a factor $\varepsilon=E_T/E_T(b)$ , assuming that the
suppression is 100\% above a threshold density  (a  parameter  in
the model), and smearing the threshold density (at the expense of
another  parameter)  the  best  fit  to  the  data  was obtained.
Extending the Blaizot's model to include fluctuations  in  number
of  NN collisions at a fixed impact parameter, NA50 data could be
fitted with a single  parameter,  the  threshold  density,  above
which  all the $J/\psi$ mesons melt \cite{ch01a}. Assumption that
all  the  $J/\psi$  mesons  melt  above  a   threshold   density,
implicitly  assume  that  QGP like environment is produced in the
collision. NA50 data  could  be  explained  in  the  conventional
approach  also,  without invoking QGP like scenario. Capella {\em
et al.} \cite{ca00} analyzed the data in  the  comover  approach.
There  also, the comover density has to be modified by the factor
$\varepsilon$. Introduction of this  adhoc  factor  $\varepsilon$
can be justified in a model based on excited nucleons represented
by strings \cite{hu00}.

Aim  of  the  present paper is to show that while in conventional
approach, nuclear suppression is not sufficient to  explain  NA50
data,  the  data  are very well described in a model of Qiu, Vary
and Zhang \cite{qiu98}, where  the  suppression  due  to  nuclear
environment is treated in an unconventional manner.

\section{model}

Recently,  Qiu,Vary  and  Zhang  \cite{qiu98} proposed a model to
describe        the        $J/\psi$        suppression         in
nucleon-nucleus/nucleus-nucleus   collisions.  For  the  sake  of
completeness, we will briefly describe the model. Qiu,  Vary  and
Zhang assumed that the production of $J/\psi$ meson is a two step
process, (i) production of $c\bar c$ pairs with relative momentum
square  $q^2$,  and  (ii)  formation  of $J/\psi$ mesons from the
$c\bar{c}$ pairs. Step (i) can be accurately calculated  in  QCD.
The  second  step,  formation  of  $J/\psi$ mesons from initially
compact  $c\bar{c}$  pairs  is  non-perturbative.  They  used   a
parametric  form  for  the  step (ii), formation of $J/\psi$ from
$c\bar{c}$ pairs. The $J/\psi$ cross section in $AB$  collisions,
at center of mass energy $\sqrt{s}$ was then written as,

\begin{equation}\label{1} \sigma_{A+B \rightarrow J/\psi + X} (s)
= K \sum_{a,b} \int dq^2 \left( \frac{\hat \sigma_{ab \rightarrow
cc}}     {Q^2}     \right)    \int    dx_F    \phi_{a/A}(x_a,Q^2)
\phi_{b/B}(x_b,Q^2) \frac{x_a x_b}{x_a + x_b} \times  F_{c\bar{c}
\rightarrow J/\psi} (q^2), \end{equation}

\noindent  where  $\sum_{a,b}$  runs over all parton flavors, and
$Q^2 = q^2 +4 m_c^2$. The  $K$  factor  takes  into  account  the
higher  order corrections. The incoming parton momentum fractions
are     fixed      by      kinematics      and      are      $x_a
=(\sqrt{x^2_F+4Q^2/s}+x_F)/2$               and              $x_b
=(\sqrt{x^2_F+4Q^2/s}-x_F)/2$.  Quark  annihilation   and   gluon
fusion  are the major sub processes for $c\bar{c}$ production. In
the leading log, they are given by \cite{be94},

\begin{eqnarray}   \label{2}  \hat  \sigma_{q\bar{q}  \rightarrow
c\bar{c}}  (Q^2)  =  &&\frac{2}{9}  \frac{4  \pi  \alpha_s}{3Q^2}
(1+\frac{\gamma}{2}) \sqrt{1-\gamma},\\
 \hat  \sigma_{gg  \rightarrow  c\bar{c}}  (Q^2)  = && \frac{ \pi
\alpha_s}{3Q^2} (1+\frac{\gamma}{2} +  \frac{\gamma^2}{16})  \log
(\frac{1+\sqrt{1-\gamma}}{1-\sqrt{1-\gamma}})                   -
(\frac{7}{4}+\frac{31}{16}\gamma)\sqrt{1-\gamma}],\\
\end{eqnarray}

\noindent  where  $\alpha_s$ is the QCD running coupling constant
and  $\gamma  =  4  m_c^2/Q^2$.  In  Eq.\ref{1}   $F_{c   \bar{c}
\rightarrow  J/\psi}(q^2)$  is  the transition probability that a
$c\bar{c}$ pair with relative momentum square $q^2$ evolve into a
physical  $J/\psi$  meson.  Qiu,  Vary  and  Zhang   \cite{qiu98}
considered   three   different   parametric  forms  (representing
different physical processes) for the transition probability. All
the three forms could describe the experimental energy dependence
of  total  $J/\psi$  cross   section   in   hadronic   collisions
\cite{qiu98}.

In  a  nucleon-nucleus/nucleus-nucleus  collision,  the  produced
$c\bar{c}$ pairs interact with nuclear medium before  they  exit.
Observed  anomalous nuclear enhancement of the momentum imbalance
in dijet production led Qiu, Vary and Zhang \cite{qiu98} to argue
that  the  interaction  of  a  $c\bar{c}$   pair   with   nuclear
environment,  increases  the  square  of  the  relative  momentum
between the $c\bar{c}$ pair. As a result some of  the  $c\bar{c}$
pairs  might  gain  enough  relative momentum squared $q^2$ to be
pushed over the  the  threshold  to  become  open  charm  mesons.
Consequently,  the  cross  sections  for  $J/\psi$ production are
reduced in comparison with  nucleon-nucleon  collisions.  If  the
$J/\psi$  meson  travel  a distance L, the transition probability
$F_{c \bar{c}} (q^2)$ in eq.\ref{1} will be changed to

\begin{equation}  \label{3}  F_{c\bar{c} \rightarrow J/\psi}(q^2)
\rightarrow F_{c\bar{c} \rightarrow J/\psi} (q^2 +  \varepsilon^2
L), \end{equation}

\noindent  with  $\varepsilon^2$  being  the  square  of relative
momentum gained by the $c\bar{c}$ pair per unit length of nuclear
medium. Of the three different parametric forms of the transition
probability, all of which fitted the  energy  dependence  of  the
$J/\psi$  cross  section  in  hadron-nucleus collisions, only the
following form,

\begin{equation} \label{4} F_{c \bar{c} \rightarrow J/\psi} (q^2)
=  N_{J/\psi} \theta(q^2) (1 - \frac{q^2}{{4m^\prime}^2 - 4 m_c^2
})^{\alpha_F}, \end{equation}

\noindent  could  describe  the  experimental  $J/\psi$ data as a
function  of   effective   nuclear   length   \cite{qiu98}.   For
completeness  purpose, we have redone the calculation of $J/\psi$
production as a function of effective  nuclear  length.  We  have
used  the  CTEQ5  parton  distribution functions \cite{cteq5}. In
Fig.1, NA50  data  \cite{ab97}  on  $J/\psi$  cross  section  for
proton-nucleon,  proton-nucleus and nucleus-nucleus collisions as
a function of the  effective  nuclear  medium  length  L(A,B)  is
shown.  The  solid  line  is  a  fit  obtained  in the model. The
parameter  values,  $KN_{J/\psi}$=.458,  $\varepsilon^2$=.225   $
GeV^2/fm$ and $\alpha_F$=1, are very close to the values obtained
in  Ref.\cite{qiu98}.  In  the  next  section,  we will use these
parameters to analyze the NA50  data  on  the  transverse  energy
dependence of $J/\psi$ to Drell-Yan ratio.

\section{$E_T$ dependence of $J/\psi$ in Pb+Pb collisions}

NA50  collaboration  presented  transverse  energy  dependence of
$J/\psi$ to Drell-Yan ratio in Pb+Pb collisions  \cite{na50}.  As
mentioned in the beginning, the data shows anomalous suppression,
which  goes  beyond  the conventional nuclear suppression. In the
present section, it will  be  shown  that,  the  data  are  fully
explained  in  the  model  of  Qiu,  Vary  and Zhang, which treat
$J/\psi$ suppression in nuclear environment in an  unconventional
manner.

The  Drell-Yan  pairs  do not suffer final state interactions and
the cross section at an impact parameter ${\bf b }$ as a function
of $E_T$ can be written as,

\begin{equation}    \label{5}    d^2\sigma^{DY}/dE_T    d^2b    =
\sigma_{NN}^{DY} \int d^2s  T_A({\bf  s})  T_B({\bf  s}-{\bf  b})
P(b,E_T), \end{equation}

\noindent where $\sigma_{NN}^{DY}$ is the Drell-Yan cross section
in  $NN$  collisions. All the nuclear information is contained in
the nuclear thickness function,  $T_{A,B}({\bf  s})  (=  \int  dz
\rho_{A,B}({\bf  s},z)$.  Presently  we  have  used the following
parametric form for $\rho_A(r)$ \cite{bl00},

\begin{equation}                                        \label{6}
\rho_A(r)=\frac{\rho_0}{1+exp(\frac{r-r_0}{a})} \end{equation}

\noindent with $a=0.53 fm$, $r_0=1.1A^{1/3}$. The central density
is  obtained  from  $\int  \rho_A(r)  d^3r  =  A$. In Eq.\ref{5},
$P(b,E_T)$ is the  probability  to  obtain  $E_T$  at  an  impact
parameter  $b$.  Geometric  model  has  been  quite successful in
explaining  the  transverse  energy  as  well   as   multiplicity
distributions  in  $AA$  collisions  \cite{ch90,ch93}. Transverse
energy distribution in Pb+Pb collisions also could  be  described
in  this model \cite{ch01a}. In this model, $E_T$ distribution is
written in terms of $E_T$ distribution in NN collisions. One also
assume that the Gama distribution, with parameters  $\alpha$  and
$\beta$  describe the $E_T$ distributions in NN collisions. Pb+Pb
data on $E_T$ distribution could be fitted with $\alpha =3.46 \pm
0.19$ and $\beta = 0.379 \pm 0.021$ \cite{ch01a}.

While  Drell-Yan  pairs  do  not suffer interactions with nuclear
matter, the $J/\psi$ mesons do. In the model  of  Qiu,  Vary  and
Zhang  \cite{qiu98},  suppression  factor  depend  on  the length
traversed  by  the   $c\bar{c}$   mesons   in   nuclear   medium.
Consequently,  we  write  the $J/\psi$ cross section at an impact
parameter ${\bf b}$ as,

\begin{equation}   \label{7}   d^2\sigma^{J/\psi}/dE_T   d^2b   =
\sigma_{NN}^{J/\psi}  \int  d^2s  T_A({\bf  s})  T_B({\bf   s-b})
S(L({\bf b,s})) P(b,E_T), \end{equation}

\noindent  where  $\sigma_{NN}^{J/\psi}$  is  the  $J/\psi$ cross
section  in  $NN$  collisions  and  $S(L({\bf  b,s}))$   is   the
suppression  factor  due to passage through a length L in nuclear
environment. At an impact parameter ${\bf b}$ and at point  ${\bf
s}$, the transverse density can be calculated as,

\begin{equation}  \label{8}  n({\bf  b,s})  =  T_A({\bf  s}) [1 -
e^{-\sigma_{NN}  T_B({\bf  b-s})}]  +   T_B({\bf   b-s})   [1   -
e^{-\sigma_{NN} T_A({\bf s})}], \end{equation}

\noindent  and  the length $L({\bf b,s})$ that the $J/\psi$ meson
will traverse can be obtained as,

\begin{equation}  \label{9}  L({\bf  b,s})=n({\bf  b,s})/2 \rho_0
\end{equation}

Suppression  factor  $S(L({\bf  b,s})$  can  be  calculated using
Eq.\ref{1}, noting that  $c\bar{c}$  pairs  gain  $\varepsilon^2$
momentum  per unit length L. Parametric value of $\varepsilon^2$,
as shown before  was  obtained  by  fitting  nucleon-nucleus  and
nucleus-nucleus $J/\psi$ cross section data containing all $E_T$.
However,   Eq.\ref{7}   corresponds   to   a   particular  $E_T$.
Accordingly, momentum gain factor $\varepsilon^2$,  needs  to  be
modified, We modify the momentum gain factor $\epsilon^2$ to take
into account the $E_T$ dependence as,

\begin{equation}   \label{10}  \varepsilon^2(E_T)=\varepsilon^2_0
\frac{L(E_T)}{\int dE_T L(E_T)} , \end{equation}

\noindent where $\varepsilon^2_0$ is the momentum gain factor for
all  $E_T$  (which  was  obtained  by fitting experimental data).
$L(E_T)$ is the  length  through  which  a  $J/\psi$  meson  with
transverse  energy  $E_T$ will travel. The length $L(E_T)$ can be
calculated \cite{vo99},

\begin{equation}  L(E_T)=  \frac{  \int  d^2b  d^2s  T_A({\bf s})
T_B({\bf  b-s})  [T_A({\bf   s})+T_B({\bf   b-s})]   P(b,E_T)   }
{2\sigma_{NN}  \rho_0  \int d^2b d^2s T_A({\bf s}) T_B({\bf b-s})
P(b,E_T)} \end{equation}

Fluctuations  of  transverse  energy  at a fixed impact parameter
plays an important role in the  explanation  of  the  NA50  data.
Above  100  GeV, i.e., approximately at the position of the knee,
the 2nd drop in the data is due to the fluctuations in $E_T$ . In
order to account for the fluctuations, following  Capella  et  al
\cite{ca00}, we calculate,

\begin{equation} F(E_T) = E_T /E_T^{NF}(E_T), \end{equation}

\noindent where, \begin{equation} E_T^{NF}(E_T) = \frac{\int d^2b
E_T^{NF}(b) P(b,E_T)}
		{\int d^2b P(b,E_T)}
\end{equation}

The   function   $F(E_T)$   is  unity  up  to  the  knee  of  the
distribution,  and   increases   thereafter,   precisely,   where
fluctuations dominates. The replacement,

\begin{equation}  L({\bf  b,s})  \rightarrow L({\bf b,s}) F(E_T),
\end{equation}

\noindent  then  properly  accounts  for  the fluctuations in the
$E_T$ distributions.

 In Fig.2, we have compared the $E_T$ distribution of $J/\psi$ to
Drell-Yan ratio, obtained in the model with the experimental data
obtained  by NA50 collaboration. In the calculation, we have used
$\sigma_{NN}$    =32    mb    and     $\sigma^{J/\psi}_{NN}     /
\sigma^{DY}_{NN}$   =53.5   \cite{bl00}.   We   obtain  excellent
agreement  with  data.  The  second  drop  at  $E_T$=100  GeV  is
correctly  reproduced.  It  may  be  emphasized  that the present
calculation is essentially a parameter free calculation. The  few
parameters of the model were obtained previously from the fitting
the  total  $J/\psi$  cross  section  in  pA  and  AA collisions.
Excellent agreement with data indicate  that  the  NA50  data  is
fully explained in terms of suppression in nuclear environment.

\section{Prediction for RHIC energy}

Present model can be used to predict $E_T$ dependence of $J/\psi$
to  Drell-Yan  ratio  at  RHIC  energy.  Recent PHOBOS experiment
\cite{phobos}  showed  that   for   central   collisions,   total
multiplicity  is larger by 70\% at RHIC than at SPS. $E_T$ can be
assumed to be increased by the same factor.  Accordingly,  scaled
the  $E_T$  distribution  for  Pb+Pb collisions at SPS energy can
represent the experimental $E_T$ distribution at RHIC energy  for
Au+Au  collisions (small mass difference between Au and Pb can be
neglected). We have fitted the rescaled $E_T$ distribution in the
geometric model  to  obtain  the  parameters,  $\alpha$=3.09  and
$\beta$=0.495
 \cite{ch01b}.    Nucleon-nucleon    inelastic    cross   section
($\sigma_{NN}$) was assumed to be 41 mb at RHIC, instead of 32 mb
at SPS \cite{bl01}.

At RHIC energy the so-called hard component which is proportional
to   number   of   binary   collisions  appear.  Model  dependent
calculations indicate that the hard component grows from 22\%  to
37\%  as  the  energy  changes  from $\sqrt{s}$=56 GeV to 130 GeV
\cite{kh01}. $J/\psi$ suppression will  strongly  depend  on  the
hard  component,  as  it effectively increases the density of the
nuclear medium. For $f$ fraction of hard  scattering,  transverse
density $n({\bf b,s})$ in Eq.\ref{8} is modified to \cite{bl01},

\begin{equation}  n_{mod}  ({\bf  b,s})  \rightarrow (1-f) n({\bf
b,s}) + f n^{hard}({\bf b,s}) , \end{equation}

\noindent  with  $n^{hard}({\bf  b,s})=\sigma_{NN}  T_A({\bf  s})
T_B({\bf b-s})$.  With  hard  component,  transverse  density  is
increased,  as  a  result,  suppression will be increased at RHIC
energy. In Fig.2, the thick solid  line  is  the  prediction  for
$J/\psi$ to Drell-Yan ratio at RHIC energy, for Au+Au collisions,
obtained with 37\% hard scattering component in the density. Very
large  suppression  is  obtained. Effect of $E_T$ fluctuations is
not visible anymore (very  large  suppression  washes  out  $E_T$
fluctuations).   It   is   interesting  to  compare  the  present
prediction with other model  calculations.  In  fig.2,  the  thin
dotted  line  is  the prediction obtained by Blaizot {\em et al.}
\cite{bl01} in model where all the $J/\psi$ mesons melts above  a
threshold  density,  essentially  in  a deconfined scenario. Very
close agreement between the predictions  obtained  in  a  nuclear
environment and in a deconfined scenario is interesting. It seems
that  it  may  not be possible to confirm the deconfinement phase
transition, which is expected to occur at RHIC energy,  from  the
$J/\psi$  data.  Recently  several  authors have proposed that at
RHIC  energy,  in  a  deconfined   scenario,   recombination   of
$c\bar{c}$  pairs  will lead to enhancement of $J/\psi$'s, rather
than its suppression \cite{recom}. Inclusion of recombination effects may mask
the  large  suppression  obtained  by  Blaizot   {\em   et   al.}
\cite{bl01}. However, nuclear suppression as calculated presently
will remain unaltered. It may then be possible to distinguish the
deconfinement phase transition from the $J/\psi$ data.

\section{Summary}
	To summarize, we have analyzed the NA50 data on transverse
energy  distribution  of  $J/\psi$  to  Drell-Yan  ratio in Pb+Pb
collisions. The data were analyzed in a model, where  suppression
of  $J/\psi$  is  due  to  gain  in  relative  square momentum of
$c\bar{c}$ pairs as it travels through the  nuclear  environment.
Some  of  the  $c\bar{c}$ pairs can gain enough momentum to cross
the threshold to become open charm mesons. The model, without any
free parameters can  well  explain  the  NA50  data  on  $J/\psi$
suppression.  Present  analysis  clearly  shows  that  it  is not
essential to assume a deconfined scenario  to  explain  the  NA50
data.  The  model  was  used  to  predict  $E_T$  distribution of
$J/\psi$  to  Drell-Yan  ratio  at  RHIC  energy.  At  RHIC  hard
component  of scattering may be important. Very large suppression
is obtained if the hard component is included. Interestingly, the
prediction obtained in the model, with only  nuclear  suppression
agrees  closely  with  the  prediction  obtained  in a deconfined
scenario. However, as suggested by several authors, recombination
of  $c{\bar  c}$  in  a  deconfinement  scenario  may   lead   to
enhancement,  rather than suppression of $J/\psi$ at RHIC energy.
Recombination effect will not  affect  the  nuclear  suppression.
Observation  of  enhanced production $J/\psi$ at RHIC energy will
then confirm deconfinement phase transition.

\begin{figure}[h]
\centerline{\psfig{figure=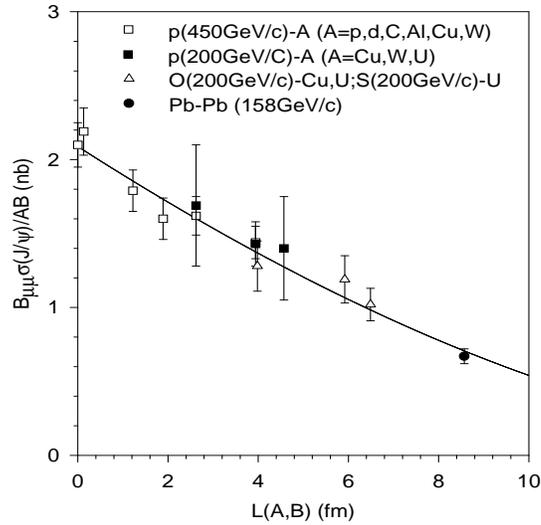,height=10cm,width=8cm}}
\vspace{-1cm}  \caption{Total  $J/\psi$  cross  sections with the
branching ratio to $\mu^+\mu^-$ in proton-nucleus, proton-nucleus
and nucleus-nucleus collisions, as a function  of  the  effective
nuclear length L(A,B).} \end{figure}

\begin{figure}[h]
\centerline{\psfig{figure=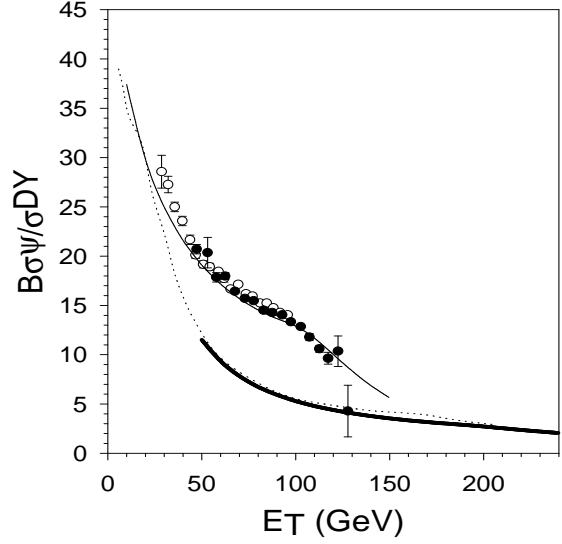,height=10cm,width=8cm}}
\vspace{-1cm}  \caption{Open  and closed circles are the $J/\psi$
to  Drell-Yan  ratio  in  a  Pb+Pb  collision  obtained  by  NA50
collaboration  in  1996 and 1998 respectively. The thin line is a
fit to the data in the model described in  the  text.  The  thick
solid  line  is  the  prediction obtained for Au+Au collisions at
RHIC energy, with 37\% hard scattering component (see text).  The
thin  dotted  line  is the prediction obtained by Blaizot {\em et
al.} {\protect \cite{bl01}} for Au+Au collisions at RHIC  energy,
in  a model where all the $J/\psi$ mesons melts above a threshold
density. } \end{figure} \end{document}